\documentclass[final,5p,twocoloumn]{elsarticle}
\usepackage[english]{babel}
\usepackage{indentfirst}
\usepackage{epsfig}
\usepackage{amsfonts,amssymb,latexsym,amsmath,enumerate}
\usepackage{bm}
\def\rot{{\rm rot}}
\def\div{{\rm div}}
\def\grad{{\rm grad}}

\journal{Physics Letters A}

\begin{document}

\begin{frontmatter}


\title{Generalized surface current method\\ in the macroscopic theory of diffraction radiation}


\author[]{D.V. Karlovets\corref{1}}
\ead{karlovets@tpu.ru}

\author[]{A.P. Potylitsyn}
\address{Tomsk Polytechnic University, Lenina ave. 30, Tomsk, 634050, Russian Federation}
\cortext[1]{Tel.: +7 3822 418906; fax: +7 3822 418901.}

\begin{abstract}
The surface current method known in the theory of electromagnetic waves diffraction is generalized to be applied for the problems of diffraction radiation 
generated by a charged particle moving nearby an ideally-conducting screen in vacuum. An expression for induced surface current density leading to the exact results 
in the theory of transition radiation is derived, and by using this expression several exact solutions of diffraction radiation problems are found. 
Limits of applicability for the earlier known models based on the surface current conception are indicated. 
Properties of radiation from a semi-plane and from a slit in cylinder are investigated at the various distances to observer.
\end{abstract}


\end{frontmatter}


\section{Introduction}
\label{Sect0}

There are several methods used in the theory of electromagnetic waves diffraction on the ideally-conducting surfaces. 
But only one of them has a clear physical interpretation: the exact method developed by V.A. Fock, where a scattered field is represented as a field of the surface current induced by an incident wave 
\cite{Fock-JETP, Fock, Vainshtein}. All other methods, such as vector Kirchhoff's integral \cite{Jakson}, double current sheet method \cite{Smythe, Smite}, or the one based on the so-called equivalence principles \cite{Schelkunoff}, don't have advantage of the clear physical sense. Actually, the first one leads to exact solution 
only introducing the rather artificial contour currents \cite{Stratton}, while the others rely upon a dual representation where magnetic currents are used \cite{Smythe, Smite, Schelkunoff, Stratton}. 
Recently, we have pointed out that in the macroscopic theory of diffraction radiation (DR) and transition radiation (TR) 
generated by the charged particles the ordinary surface current method turns out to be approximate \cite{JETP-2}. 
In contrast to the classical diffraction theory, this method provides the right result only within some simplest cases. 
It means, that solutions for DR found with the use of this method turn out approximate too. 
For example, the well-known solution for DR of a particle moving nearby an ideally-conducting semi-plane \cite{Kazantsev} has been shown to have some limits of applicability. 

In this paper, the surface current method is generalized for problems of TR and DR generated by a charged particle on an ideally-conducting surface.
The field of a scattered wave is represented as a radiation field of the surface current formed by the dipoles disposed on an ideally-conducting surface. 
This representation allows to explain the widely-used dual formalism (e.g. in works \cite{Smythe, Smite, Schelkunoff, Stratton, Shkvarunets}). 
Particularly, the use of magnetic currents formed by the ``true'' (Dirac) magnetic dipoles is physically possible due to equivalence of such a dipole to an ordinary magnetic dipole 
in vacuum \cite{Vainshtein, Frank-2}. It is shown that in order to get the right result for TR and DR, it is necessary to have a non-zero component of the surface current density 
which is normal to the screen. In the special case of plane-wave diffraction, the method developed leads to the well-known integral equation for the surface current density \cite{Fock-JETP, Fock, Cullen}. In the case of TR studied at the frequencies lower than the optical ones, the method leads to the well-known results by Korkhmazyan and Pafomov \cite{Korhmazyan-incident, Korhmazyan, Pafomov}. Finally, the method allows to find the exact solutions in the problems of diffraction radiation and Smith-Purcell radiation for the wide region of parameters.

\section{Integral equations for the surface currents}
\label{Sect1}

The exact macroscopic theory of plane waves diffraction is based on the use of the well-known integral equation for the surface current density induced by an incident field \cite{Fock-JETP, Fock, Cullen}. This method is also used to consider TR, arising when a charged particle crosses an ideally-conducting screen \cite{Tilinin, Suterlin}. 
However, if the incident field satisfies the inhomogeneous Maxwell's equations (e.g. the own field of a particle), this method turns out to be approximate \cite{JETP-2}. 
This rather unexpected fact may be explained taking into account the derivation of this equation using the so-called double-current layer formulation \cite{Tamm}. 
On the other hand, this formulation allows to find the new integral equations determining a surface current density in the case when the incident field is not a plane wave.
Solutions of these generalized equations allow to find the surface current density, and therefore to find an exact solution of a problem.

Physically, when a field (no matter what its nature) falls on a screen with unit normal $\bold n$ it induces a dipole moment resulting in appearance of an additional field. 
In other words, the scattered wave may be represented as a field of the surface current formed by the dipoles distributed on a screen.
Such representation corresponds to the ordinary dipole approximation in the microscopic theory of diffraction radiation \cite{Tishchenko}.
The surface distribution of electric dipole moment is known as a double sheet (layer) \cite{Tamm}, that is a surface where the density of surface charges 
$\rho_e \propto \bold n \cdot  \bold E$ changes its sign but preserves its absolute value. 
The corresponding Maxwell's equations for the double surface current density are written as:
\begin{eqnarray}
\bold E (\bold r, \omega) = \displaystyle \frac{i c}{\omega} \Big ( \grad \ \div + \frac{\omega^2}{c^2} \Big ) \bold A,
\bold H (\bold r, \omega) = \rot \bold A,\cr \bold A = \displaystyle \frac{1}{c} \int \bold j_s^e ({\bold r}^{\prime}, \omega )
\frac{e^{i \omega |\bold r - {\bold r}^{\prime}| / c}}{|\bold r - {\bold r}^{\prime}|} d S^{\prime}, \bold j_s^e = \frac{c}{2 \pi}\bold n \times \bold H 
\label{eq1}
\end{eqnarray}
where the dependence of current in the right-hand side upon the field in the left-hand side makes them integral equations for fields.

There also another method may be used, where a sign of the surface electric charges doesn't change when crossing the screen, but a sign of magnetic charges 
$\rho_m \propto \bold n \cdot  \bold H$ changes. This representation leads to the dual Maxwell's equations for the double magnetic surface current density:
\begin{eqnarray}
\bold H (\bold r, \omega) = \displaystyle \frac{i c}{\omega} \Big ( \grad \ \div + \frac{\omega^2}{c^2} \Big ) \bold {\tilde A},
\bold E (\bold r, \omega) = - \rot \bold {\tilde A}, \cr \bold {\tilde A} = \displaystyle \frac{1}{c} \int \bold j_s^m ({\bold r}^{\prime}, \omega )
\frac{e^{i \omega |\bold r - {\bold r}^{\prime}| / c}}{|\bold r - {\bold r}^{\prime}|} d S^{\prime}, \bold j_s^m = - \frac{c}{2 \pi}\bold n \times \bold E 
\label{eq2}
\end{eqnarray}
and corresponds to a double magnetic sheet, where the ``current density'' (axial vector) $\bold j_s^m$ is formed by magnetic dipoles. 
The problem of a plane wave diffraction on a screen with permittivity $\epsilon = \infty$ is known to be equivalent to the problem of diffraction 
on a complementary screen with $\mu = \infty$ \cite{Landau-8}. It is the last case that corresponds to the dual representation.
As will be shown in Sect.\ref{Sect2}, a ``true'' magnetic dipole (a pair of magnetic charges) is completely equivalent in vacuum to an ordinary magnetic dipole 
(elemental current loop) that only makes the sense to use the dual formalism.

The fields in the right-hand sides of Eqs.(\ref{eq1}),(\ref{eq2}) consist of the incident fields $\bold H_i, \bold E_i$ and the scattered ones $\bold H_s, \bold E_s$, 
while the fields in the left-hand side are commonly observed at the far distances where only the scattered field has place (wave zone).
Note that within the volume bounded by the screen, the total fields $\bold H = \bold H_i, + \bold H_s,\ \bold E = \bold E_i + \bold E_s$ satisfy the homogeneous Maxwell's equations. 
Substituting the expression for magnetic field in the one for the surface current in Eq.(\ref{eq1}), we can write down the ordinary Fock's integral equation 
for the current density \cite{Fock-JETP, Fock, Cullen}:
\begin{eqnarray}
& \displaystyle \bold j_s^e ({\bold r}^{\prime},\omega ) = \displaystyle \frac{c}{2 \pi} \bold n \times \bold H_i - \cr & - \displaystyle \frac{1}{2 \pi} \bold n \times 
\int \bold j_s^e({\bold r}^{\prime \prime}, \omega ) \times \grad \displaystyle \frac{e^{i \omega |{\bold r}^{\prime} - {\bold r}^{\prime \prime}| / c}}{|{\bold r}^{\prime} 
- {\bold r}^{\prime \prime}|} d S^{\prime \prime}. \label{FockEq.}
\end{eqnarray}
From Eqs.(\ref{eq2}), the similar integral equation for magnetic current density follows:
\begin{eqnarray}
& \displaystyle \bold j_s^m ({\bold r}^{\prime},\omega ) = - \displaystyle \frac{c}{2 \pi} \bold n \times \bold E_i - \cr & - \displaystyle \frac{1}{2 \pi} \bold n \times 
\int \bold j_s^m({\bold r}^{\prime \prime}, \omega ) \times \grad \displaystyle \frac{e^{i \omega |{\bold r}^{\prime} - {\bold r}^{\prime \prime}| / c}}{|{\bold r}^{\prime} 
- {\bold r}^{\prime \prime}|} d S^{\prime \prime}. \label{DualFockEq.}
\end{eqnarray}
Because of the boundary conditions for ideal conductor, integration in the last equation is performed along the aperture, while it is performed along the screen in Eq.(\ref{FockEq.}) (see Fig.1).
It is this separation that allows not to use the well-known Kirchhoff's approximation \cite{Vainshtein, Stratton}. 
That is why, the method based on solution of these equations commonly leads to exact solution of the diffraction problem.
\begin{figure}
\includegraphics[width=7.00cm, height=6.00cm]{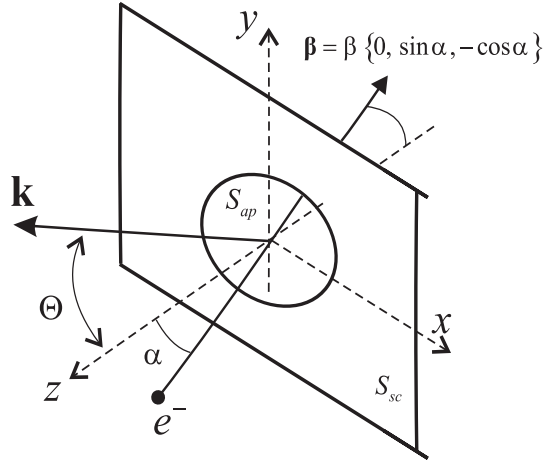}
\caption{\label{Fig1}Generation of diffraction radiation by passing of a charged particle through an aperture in the screen.
Transition radiation appears when the aperture radius approaches zero }
\end{figure}

Eqs.(\ref{FockEq.}),(\ref{DualFockEq.}) may be directly used for the plane-wave diffraction only \cite{JETP-2}, but they may be generalized for the case 
when the incident wave satisfies the \textit{in}homogeneous Maxwell's equations. In this case, one should replace the fields $\bold E, \bold H$ in (\ref{eq1}),(\ref{eq2}) by the radiation fields 
that are the difference between the total fields and those of the sources: 
\begin{eqnarray}
\displaystyle && \bold E \rightarrow \bold E^R = \bold E - \bold E^0, \ \bold H \rightarrow \bold H^R = \bold H - \bold H^0, \label{Replacement}
\end{eqnarray}
Note that the surface currents are induced by the total fields $\bold E$ and $\bold H$, as before.
By using these considerations and Eqs.(\ref{eq1}),(\ref{eq2}) written for the radiation fields, one can derive the new integral equations for a problem with external sources:
\begin{eqnarray}
& \displaystyle \bold j_s^e ({\bold r}^{\prime},\omega ) = \frac{c}{2 \pi} \bold n \times \bold H^0 - \cr & - \displaystyle \frac{1}{2 \pi} \bold n \times \int \bold j_s^e({\bold r}^{\prime \prime}, \omega ) \times \grad \displaystyle \frac{e^{i \omega |{\bold r}^{\prime} - {\bold r}^{\prime \prime}| / c}}{|{\bold r}^{\prime} - {\bold r}^{\prime \prime}|} d S^{\prime \prime} + \cr & + \displaystyle \frac{c}{(2 \pi )^2} \bold n \times \int [\bold n \times \bold H^0] \times \grad \displaystyle \frac{e^{i \omega |{\bold r}^{\prime} - {\bold r}^{\prime \prime}| / c}}{|{\bold r}^{\prime} 
- {\bold r}^{\prime \prime}|} d S^{\prime \prime}, \cr & \displaystyle \bold j_s^m ({\bold r}^{\prime},\omega ) = - \displaystyle \frac{c}{2 \pi} \bold n \times \bold E^0 - \cr & - 
\displaystyle \frac{1}{2 \pi} \bold n \times \int \bold j_s^m({\bold r}^{\prime \prime}, \omega ) \times \grad \frac{e^{i \omega |{\bold r}^{\prime} - {\bold r}^{\prime \prime}| / c}}{|{\bold r}^{\prime} - {\bold r}^{\prime \prime}|} d S^{\prime \prime} - \cr & - \displaystyle \frac{c}{(2 \pi )^2} \bold n \times \int [\bold n \times \bold E^0] \times \grad \displaystyle \frac{e^{i \omega |{\bold r}^{\prime} - {\bold r}^{\prime \prime}| / c}}{|{\bold r}^{\prime} - {\bold r}^{\prime \prime}|} d S^{\prime \prime}. \label{GenFockEqs.}
\end{eqnarray}

Let us discuss the common and opposite features of the integral equations for the plane waves (\ref{FockEq.}),(\ref{DualFockEq.}) and the generalized ones for the fields of external sources 
(\ref{GenFockEqs.}). The ordinary Fock's equation is usually solved via iteration method, where a parameter of expansion represents the ratio of the wavelength to the radius of the surface curvature 
\cite{Fock-JETP, Fock}. Therefore, the corresponding dual equation derived may be solved using the very same technique. It allows to generalize the dual method used by Smythe, Schelkunoff et al. 
(see e.g. \cite{Smythe, Smite, Schelkunoff, Stratton}) for the case of a surface with arbitrary curvature. At the same time, the exact solutions of Eqs.(\ref{GenFockEqs.}) 
may be found in the compact universal form suitable even for a concave surface. By using the same iteration technique, one can find:
\begin{eqnarray}
\displaystyle \bold j_s^e({\bold r}^{\prime},\omega ) = \frac{c}{2 \pi} \bold n \times \bold H^0, \ \bold j_s^m({\bold r}^{\prime},\omega ) = - \frac{c}{2 \pi} \bold n \times \bold E^0. \label{Sol}
\end{eqnarray}
Formally, these exact solutions are very similar to those of Eqs.(\ref{FockEq.}),(\ref{DualFockEq.}) for a flat surface, i.e. when only the 0th  term of expansion is taken \cite{Fock-JETP, Fock, Suterlin}.
It might be explained in the following way. The high-order terms in solution of Eq.(\ref{FockEq.}) take into account the secondary reflections of an incident plane wave by the concave surface.
But if the initial field is not the one of a plane wave, the secondary reflections occur \textit{not} for the field of an external source, but for the field of the surface current radiation.
Therefore, the macroscopic theory of DR and TR based on the use of Eqs.(\ref{GenFockEqs.}) is always the ``one-reflection'' theory. 
It hampers its application for the too curved surfaces, where the secondary reflections occur. Note that not both of solutions (\ref{Sol}) lead to the exact result 
in the problem of transition radiation on an inclined screen (only magnetic current $\bold j_s^m$ does \cite{JETP-2, Shkvarunets}). This feature is discussed in Sect.\ref{Sect3}.

\section{Justification for the dual representation}
\label{Sect2}

There are two types of integral equations derived: one, where the scattered wave is generated by the surface current of electric dipoles and another one,
where magnetic current formed by the ``true'' magnetic dipoles radiates. The last formulation is commonly used (see e.g. \cite{Jakson, Smythe, Smite, Schelkunoff, Stratton, Shkvarunets} et al.),
but there are only few works where the physical validity of such method is proved. Here we produce a rather simple proof of possibility to use such a method in the theories 
of diffraction and diffraction radiation.

The dual Maxwell's equations are derived from the ordinary ones by the following substitute (see e.g. \cite{Ginzburg}):
\begin{eqnarray}
\displaystyle && \bold E \rightarrow \bold H, \ \bold H \rightarrow - \bold E, \ \varepsilon \rightarrow \mu, \label{Subs}
\end{eqnarray}
or in tensor form: 
$H^{\mu \nu} = (-\bold D, \bold H) \rightarrow \tilde{F}^{\mu \nu} = (-\bold B, - \bold E) = 1/2 \ \epsilon^{\mu \nu \eta \sigma}{F}_{\eta \sigma}$,
where $\bold D (\bold r, \omega ) = \varepsilon (\omega) \bold E (\bold r, \omega )$ - electric displacement, 
$\bold H (\bold r, \omega ) = \bold B (\bold r, \omega )/\mu(\omega)$ - macroscopic magnetic field. 
It is essential to note, that the very same substitute takes place for the radiation fields of electric and magnetic dipoles (see e.g. Ref.\cite{Bordovitsyn-correct}). 
It means, that formally a problem of radiation of an ordinary magnetic dipole (elemental current loop) 
is completely equivalent in vacuum to the problem of radiation of a ``true'' (Dirac) dipole.
In other words, the radiation fields of such dipoles are equal to each other\footnote[1]{See also discussion on this topic concerning Cherenkov radiation of dipoles in Ref.\cite{Frank-2}}. 
Let's produce the more rigorous proof of this statement.

The radiation fields of a given electric current density in the wave zone are found to be:
\begin{eqnarray}
\displaystyle && \bold H^R_e (\bold r_0, \omega ) = \frac{i  (2 \pi )^3}{c} \frac{e^{ik r_0}}{r_0} \bold k \times \bold {j}^e (\bold k, \omega), 
\cr \displaystyle && \bold E^R_e (\bold r_0, \omega ) = - \sqrt{\frac{\mu}{\varepsilon}} \bold {e}_0 \times \bold H^R_e, \label{Rad}
\end{eqnarray}
$\bold {j}^e (\bold k, \omega)$ - Fourier transform of the current density $\bold {j}^e (\bold r, t)$, $\bold k = \bold {e}_0 \sqrt{\varepsilon \mu} \ \omega/c$ - wave vector in medium.
From this, the radiation fields for magnetic current density follow with the use of (\ref{Subs}):
\begin{eqnarray}
\displaystyle && \bold E^R_m (\bold r_0, \omega ) = - \frac{i  (2 \pi )^3}{c} \frac{e^{ik r_0}}{r_0} \ \bold k \times \bold {j}^m (\bold k, \omega), \cr \displaystyle &&
\bold H^R_m (\bold r_0, \omega ) = \sqrt{\frac{\varepsilon}{\mu}} \bold {e}_0 \times \bold E^R_m. \label{RadDual}
\end{eqnarray}

The currents of both types would be equivalent if their radiation fields are equal to each other. 
Demand for equality of magnetic fields $\bold H^R_e$ and $\bold H^R_m$ leads to the following correlation of currents 
(equality of electric fields follows automatically):
\begin{eqnarray}
\displaystyle \bold {j}^e (\bold k, \omega ) = - \frac{c}{\omega} \frac{1}{\mu (\omega )} \ \bold k \times \bold {j}^m (\bold k, \omega ), \label{ElCur}
\end{eqnarray}
or in space variables:
\begin{eqnarray}
\displaystyle \bold {j}^e (\bold r, \omega ) = \frac{i c}{\omega} \frac{1}{\mu (\omega )}\ \rot \ \bold {j}^m (\bold r, \omega ). \label{ElCurSp}
\end{eqnarray}
This relation allows to find an electric current density which is equivalent to the given magnetic current density, i.e. producing the same radiation fields.
Let's take for example the current of a ``true'' magnetic dipole (a pair of magnetic charges) with moment $\bm \mu$ in the rest frame of reference:
\begin{eqnarray}
\displaystyle && \bold {j}^m (\bold r, \omega) = - i \omega \bm {\mu}(\bold r, \omega)  \ \delta (\bold r), \label{MagDip}
\end{eqnarray}
According to Eq.(\ref{ElCurSp}) the corresponding electric current is found to be:
\begin{eqnarray}
\displaystyle && \bold {j}^e (\bold r, \omega) = \frac{1}{\mu (\omega )}\ c \ \rot \ \bm {\mu}(\bold r, \omega)  \delta(\bold r). \label{ElDipCorr}
\end{eqnarray}
For non-magnetic media ($\mu (\omega) = 1$) this is exactly the current produced by an ordinary magnetic dipole moment $\bm \mu$ in the rest frame of reference 
(see e.g. \cite{Ginzburg, Bordovitsyn-correct}). Thus, the surface magnetic current used in the theories of diffraction \cite{Smythe, Smite, Schelkunoff, Stratton} and TR \cite{Shkvarunets} 
is equivalent to the ordinary electric current of magnetic dipoles. This fact may be considered just as a consequence of the theorem of equivalence mentioned \cite{Landau-8}.

\section{Surface current density induced by the particle field}
\label{Sect3}

The expression for electric current density (\ref{Sol}) doesn't lead to the well-known results for TR, 
even in the simplest case of the normal incidence of a particle with normalized energy $\gamma = E/m_0c^2$ on a flat ideally-conducting screen in vacuum \cite{JETP-2}.
Moreover, the expression for surface current density derived in the familiar paper \cite{Kazantsev} doesn't allow to get the right transition from the formulas of DR generated 
by a particle moving through a slit in an inclined screen to the formulas of TR when the slit width approaches zero. This transition has a place only in the ultrarelativistic case
\cite{Pap-NIMB98, NP-Artru}, that was only investigated before. On the other hand, the magnetic current density (\ref{Sol}) leads to exact transition between DR formulas and the ones for TR 
in a problem of the radiation on a slit \cite{JETP-2}. By using the formula for magnetic current and taking into account the results of Sect.\ref{Sect2}, 
one can derive the exact expression for electric current density. 

First of all, let us assume what the right expression for the current density might be. As it is commonly supposed to have no normal to the screen component, the vector $\bold j_s^e$ should be determined as the cross product of the unit normal (polar vector) $\bold n=\{0, 0, 1\}$ and some axial vector.
In fact, there may be only a few variants: 
\begin{eqnarray}
a.) \bold j_s^e \propto \bold n \times \bold H^0 = \bold n \times [{\bm \beta} \times \bold E^0],\label{Current1}
\end{eqnarray} 
or 
\begin{eqnarray}
b.) \bold j_s^e \propto \bold n \times [\bold n \times \bold E^0],\label{Current2}
\end{eqnarray} 
and also 
\begin{eqnarray}
c.) \bold j_s^e \propto \bold n \times [\bold e_0 \times \bold E^0],\label{Current3}
\end{eqnarray} 
where $\bold E^0, \bold H^0$ - fields of a particle, $\bm {\beta} = \bold v / c$, and let the constant be chosen as $c/2 \pi$.
Since we have no other vectors in the problem, it is impossible to construct any other appropriate expression for the current density. 

However, it turns out that none of these variants leads to the right result for backward TR, arising when an electron crosses the screen obliquely at the angle $\alpha$ to the normal.
As Fig.2 shows, the close result is observed only for the first variant in the ultrarelativistic case when the asymmetry vanishes. 
This expression ($a.)$) leads to the following radiation intensity in the simplest case $\alpha = 0$ \cite{JETP-2}:
\begin{eqnarray}
\frac{d^2 W}{d \omega d\Omega} = c \ r_{0}^2 |\bold {E}_e^R|^2 = \frac{e^2}{\pi^2 c} \frac{\beta^4 \sin^2{\Theta} \cos^2{\Theta}}{(1 - \beta^2 \cos^2{\Theta})^2},\label{Intensity}
\end{eqnarray} 
where the radiation field $\bold {E}_e^R$ depends upon the current density according to Eq.(\ref{Rad}), and $\Theta$ is the polar angle of the backward emission, see Fig.1.
Compared to well-known results, this expression has an extra term: $\beta^2 \cos^2{\Theta}$ \cite{Landau-8}. 
For the sake of the reader's convenience, we omit the rather monotonous but not difficult calculations of the radiation intensity for other variants of the current density. 
Some mathematical details may be found in Ref.\cite{JETP-2}. 

On the contrary, the choice of current density in the form $\bold j_s^e \propto \bold e_0 \times [\bold n \times \bold E^0]$ leads exactly to the result for TR 
obtained for the first time by Korkhmazyan \cite{Korhmazyan-incident, Korhmazyan, Pafomov}, as Fig.2 d.) shows. 
Unlike our preliminary supposition, this current has all three components including the one perpendicular to the screen.
This feature will be discussed hereinafter.
\begin{figure*}
\center \includegraphics[width=16.00cm, height=10.50cm]{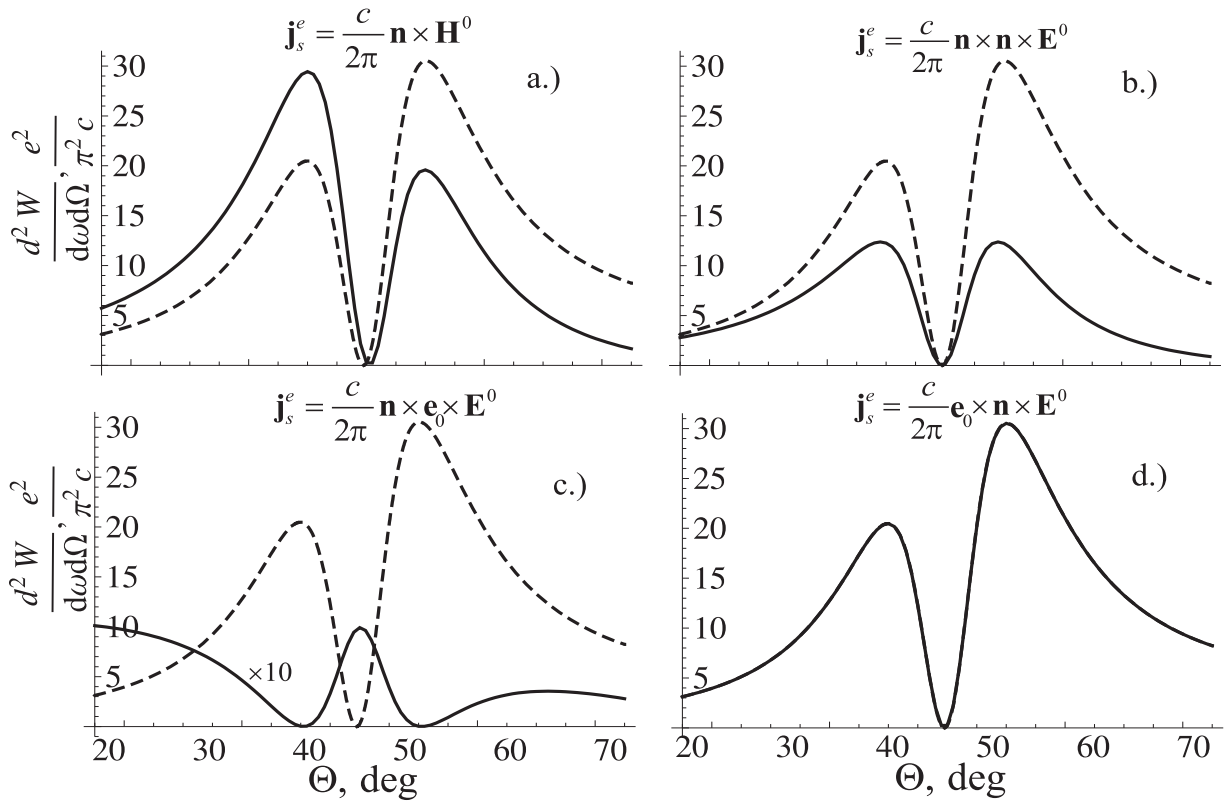}
\caption{\label{Fig2}Angular distributions of TR for the different expressions for the surface current density. 
Dashed line: calculations according to Korkhmazyan's formula \cite{JETP-2, Korhmazyan-incident, Pafomov}. 
Parameters: $\gamma = 10$, $\alpha = \pi/4$, unit radiation vector $\bold {e}_0 = \{\sin{\Theta} \sin{\Phi}, \sin{\Theta} \cos{\Phi}, \cos{\Theta}\}$, 
and $\Phi = 0$}
\end{figure*}

Another and of course more rigorous way to find the electric current density is to use the expression for magnetic current density $\bold j_s^m$ (\ref{Sol}) 
which was proved to provide the right results for TR \cite{JETP-2, Shkvarunets}, and to apply transformation rules (\ref{Subs}).
We shall search for expression of the current density suitable at the arbitrary distances to observer $r_0$.
For this purpose, one should find a solution of the wave equation for electric field in vacuum which is found to be:
\begin{eqnarray}
\displaystyle && \Big ( \Delta + \frac{\omega^2}{c^2} \Big ) \bold E^e = - i \frac{4 \pi}{\omega} \Big (\grad \ \div + \frac{\omega^2}{c^2} \Big ) \bold j^e. \label{WaveEq}
\end{eqnarray}
Convolution of the right-hand side with the infinite space Green's function allows to write down the final result for the field 
(cumbersome calculations are also omitted):
\begin{eqnarray}
& \displaystyle \bold E^e (\bold r_0, \omega ) = \frac{i}{\omega} \int \Big \{\bold {j}^e \ \Big( k^2 + \frac{ik}{|\bold r_0 - \bold r|} 
 - \frac{1}{|\bold r_0 - \bold r|^2} \Big ) \cr \displaystyle & \quad \displaystyle - \frac{\bold r_0 - \bold r}{|\bold r_0 - \bold r|}
\Big (\bold {j}^e, \frac{\bold r_0 - \bold r}{|\bold r_0 - \bold r|}\Big ) 
\Big (k^2 + \frac{3ik}{|\bold r_0 - \bold r|} \cr \displaystyle & \qquad \qquad \qquad \displaystyle - \frac{3}{|\bold r_0 - \bold r|^2} \Big) \Big \} 
\frac{e^{i\omega|\bold r_0 - \bold r|/c}}{|\bold r_0 - \bold r|}d^3 r. \label{ESol}
\end{eqnarray}
In the microscopic theory, the radiation of the particles occupied the space region of $r_{eff} \ll \lambda$ is often considered in the dipole approximation.
For this case, Eq.(\ref{ESol}) reduces to (\ref{Rad}) where $\varepsilon , \mu = 1$ should be put. 
In the macroscopic theory being considered here, the current density
\begin{eqnarray}
&& \displaystyle \bold j^e (\bold r, \omega ) = \bold j_s^e (\bold r, \omega ) \delta(z) \label{Current}
\end{eqnarray}
is induced by the field of a charged particle whose Fourier components decay exponentially. 
Hence, the effective dimensions of the surface current density $\bold j_s^e (\bold r, \omega )$ for the case of TR or DR and $\alpha \ll \pi/2$ coincide with those 
of the particle field: $r_{eff} \approx \gamma \lambda / 2\pi$ \cite{Jakson}.
In relativistic case it is much larger than a wavelength, and the radiation field at the distances $r_0 \gg \lambda/2 \pi$ is found as:
\begin{eqnarray}
\displaystyle \bold E^R_e = - \frac{i}{\omega}  \int \bold k(\bold r) \times \bold k (\bold r) \times \bold {j}_s^e (\bold r, \omega) 
\frac{e^{ik|\bold r_0 - \bold r|}}{|\bold r_0 - \bold r|} dS, \label{ERad}
\end{eqnarray}
where the wave vector is directed from the point of integration $\bold r = \{x, y, 0\}$: $\bold k(\bold r) = k (\bold r_0 - \bold r)/|\bold r_0 - \bold r|$,
but its absolute value is still equal to $k = \omega / c$.

The similar expression for magnetic current density is found by using formula (\ref{Subs}):
\begin{eqnarray}
\displaystyle \bold E^R_m = - \frac{i}{c} \int \bold k(\bold r) \times \bold {j}_s^m (\bold r, \omega) \frac{e^{ik|\bold r_0 - \bold r|}}{|\bold r_0 - \bold r|} dS. \label{ERadMag}
\end{eqnarray}
Comparison of Eqs.(\ref{ERad}), (\ref{ERadMag}) allows to find electric current density from the condition $\bold E^R_e = \bold E^R_m$. 
Since magnetic current is determined by Eq.(\ref{Sol}), the surface electric current density is found to be:
\begin{eqnarray}
\displaystyle && \bold {j}_s^e = \frac{c}{2 \pi} \frac{\bold r_0 - \bold r}{|\bold r_0 - \bold r|} \times [\bold n \times \bold E^0 ], \label{ElCurGen}
\end{eqnarray}
At the distances which are much larger than parameter $\gamma \lambda/ 2\pi$, dependence of the wave vector $\bold k (\bold r)$ upon $\bold r$ may be neglected. 
It leads to the very same formula for the current density, as was found from the general considerations:
\begin{eqnarray}
&& \bold {j}_s^e \approx \displaystyle \frac{c}{2 \pi}\ \bold e_0 \times \bold n \times \bold E^0 = \cr 
\displaystyle && \ = \frac{c}{2 \pi} \Big \{-\cos{\Theta}\ \bold E^0_x,-\cos{\Theta} \ \bold E^0_y, \ (\bold e_0, \bold E^0_{\parallel}) \Big \}, \label{ElCurWave}
\end{eqnarray}
The main feature of the expression derived is the non-zero normal component of the current $j_z \propto (\bold e_0, \bold E^0_{\parallel})$, $\bold E^0_{\parallel}= \{E^0_x, E^0_y\}$.
For the simplest case of normal incidence of a fast particle on a screen ($\alpha = 0$), it is negligibly small compared with transverse components 
for the small angles of emission $\Theta \ll 1$ that is typical for the ultrarelativistic case $\gamma \gg 1$. 
But for the more realistic geometry of oblique incidence, the backward radiation is emitted in the vicinity of the specular reflection direction, 
and the case $j_z \sim j_{\parallel}$ may be realized for the angles $\Theta - \pi/2 \ll 1$.
Dependence of the transverse current components upon the particle field $\bold j_{\parallel} \propto \bold E^0_{\parallel}$
is similar to the one in the microscopic theory of DR \cite{Tishchenko}. 
However, it is the normal component of current density $j_z$ that only results in the non-zero radiation intensity
in a plane of the screen: $\Theta = \pi/2$. Moreover, only  the presence of this component leads to the right asymmetry in the backward TR angular distributions 
for the case of oblique incidence: compare b.) and d.) in Fig.2.

The presence of a non-vanishing normal component of the surface current may be explained also in the following way.
When a field of an external source falls on the surface of a good conductor, it induces polarization currents, and the skin effect occurs.
As the conductivity goes to infinity, the skin depth goes to zero, and finally we get an ideally-conducting infinitely thin screen.
But in order to use the linear theory of skin effect appropriate for mm and sub-mm ranges of wavelengths, the skin depth should be much larger than a free length of a conduction electron.
On the other hand, the condition of applicability for macroscopic electrodynamics being used requires the skin depth to be much larger
than an effective dimension of the spatial averaging (physically infinitesimal volume) \cite{Landau-8}. 
It means that macroscopically infinitely thin screen nevertheless has a finite width, exceeding dimension of the physically infinitesimal volume, 
which is also greater than a conduction electron free length (see also discussion in Ref.\cite{Jakson}, pp. 20-21).

Thus, in order to get the right results for transition radiation of a charged particle crossing the ideally-conducting screen, 
it is necessary to have the non-zero normal component of the surface current density. 
In other words, it is impossible to construct an expression for the current density with $j_z = 0$, 
that would lead exactly to the results for TR by Ginzburg and Frank, Korkhmazyan, Pafomov et al.
It means particularly that the models for diffraction radiation and Smith-Purcell radiation
based upon the use of a surface current with zero normal component turn out approximate: \cite{Kazantsev, Tilinin, Suterlin, Bolotovsky-66, Sedrakyan, Brownell, Trotz, Kube-2} et al.
The region of their validity is determined by the inequality:
\begin{eqnarray}
&& \displaystyle j_z \ll |\bold j_{\parallel}|. \label{InEq}
\end{eqnarray}
For example, for TR and DR it is the case when the angle of incidence is small enough: $\alpha \ll \pi/2$,
and the particle energy is sufficiently large: $\gamma \gg 1$. 

We would like to emphasize that any surface current model developed for DR
should be at first tested in a problem where the exact solution is well-known and experimentally verified,
such as the problem of transition radiation having the same physical origin.
After that, it is possible to apply such a model to the more complicated geometries.
This verification was done neither in the works cited nor in the other papers devoted to the macroscopic theory of DR and Smith-Purcell radiation.
On the contrary, the expression for surface current (\ref{ElCurWave}) leads exactly to the well-known formulas of TR 
(e.g. \cite{Landau-8, Korhmazyan-incident, Korhmazyan, Pafomov} et al.), that is why it is naturally to use this expression for the problems of DR.

In conclusion of the paragraph, let us discuss the features of the radiation field generated by the surface current (\ref{ElCurGen}) at the various distances to observer $r_0$. 
Firstly, the expression for radiation field (\ref{ERad}) may be written in the form:
\begin{eqnarray}
\displaystyle && \bold E^R = \frac{i \omega}{c^2} \int \bold {j}_s^e (\bold r, \omega) 
\frac{e^{ik|\bold r_0 - \bold r|}}{|\bold r_0 - \bold r|} dS. \label{ERadSim}
\end{eqnarray}
At the large distances $r_0 \gg r_{eff} \approx \gamma \lambda / 2\pi$ one may make an analog of the standard multipole expansion in the field amplitude.
With an accuracy up to the first powers of $r/r_0$ it leads to the following formula:
\begin{eqnarray}
&& \displaystyle \bold E^R \simeq \frac{i \omega}{2 \pi c r_0} \int \Big ( \bold e_0 \times \bold n \times \bold {E}^0 \Big (1 + 2 \frac{(\bold r, \bold r_0)}{r_0^2} \Big ) 
\cr && \qquad \qquad \qquad \qquad \displaystyle - \bold n \frac{(\bold r, \bold {E}^0)}{r_0}\Big ) e^{ik|\bold r_0 - \bold r|} dS, \label{Expansion}
\end{eqnarray}
As one can see, it describes the spherical waves propagating from the origin of coordinates.
However the finite dimensions of the radiation source ($\gamma \lambda \gg \lambda$ with $\gamma \gg 1$)
reveal in the fact that the radiation field becomes ``non-transverse'': $(\bold E^R, \bold e_0) \ne 0$. 
It takes place not due to the static field as in the near-field: $r_0 \precsim \lambda/2 \pi$, but because of the fact
that the waves emitted from the screen surface reach the observation point with different directions of the wave vectors.
Note that at the distances $r_0 \precsim \gamma \lambda / 2\pi$ the field cannot be expanded into $r/r_0$ series.
In this case, it doesn't represent the superposition of the spherical waves, but it is still the radiation field as long as the condition $r_0 \gg \lambda/2 \pi$
is fulfilled.

\section{DR and TR of a particle on a semi-plane}
\label{Sect4}

The expression for the surface current density derived (\ref{ElCurGen}) allows to find the exact solutions of diffraction radiation problems.
The simplest one is the radiation arising when a particle moves close to an ideally-conducting semi-plane.
As was shown recently, the major difference between solution obtained in the familiar paper \cite{Kazantsev} 
and the one based on the double-layer formulation (i.e. via the generalized surface current method) consists in the fact 
that the radiation intensity doesn't vanish in the plane of the screen according to the model developed \cite{JETP-2}.
One can see that this difference appears due to neglect by the normal component of the surface current density $j_z$ in the work by Kazantsev and Surdutovich.
This component is responsible for TR at the large polar angles $\Theta - \pi/2 \ll 1$, so the region of validity for solution \cite{Kazantsev}
is restricted within the small values of emission angle: $\Theta \ll 1$ and the large values of the particle energy $\gamma \gg 1$. 
On the other hand, if the particle moves parallel to a semi-plane (or a grating, as in geometry of Smith-Purcell radiation), 
the error of solution \cite{Kazantsev} noticeably increases in the region of the small angles of emission even in the ultrarelativistic case (close to the plane of a grating).
\begin{figure}
\center \includegraphics[width=3.30cm, height=5.00cm]{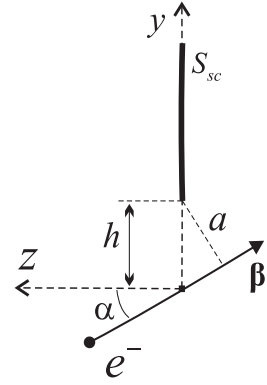}
\caption{\label{Fig3}Generation of diffraction radiation by passing of a charged particle at the distance $a = h \cos{\alpha}$ to the edge of a semi-plane}
\end{figure}

Consider a particle moving nearby an edge of a screen, and the angle between its trajectory and the screen normal is $\alpha$, see Fig. 3.
Analytical expression for the field of diffraction radiation is found as a field of the surface current (\ref{ElCurGen}) in the wave zone \cite{JETP-2}:
\begin{eqnarray}
\displaystyle \bold {E}^{DR}_h = \frac{e}{\pi c} \frac{e^{i k r_0}}{r_0} \frac{1/2}{\sqrt{1 + (\beta \gamma e_x)^2} } \Big (\frac{\cos{\alpha}}{\gamma} \sqrt{1 + (\beta \gamma e_x)^2} \cr 
\displaystyle - i (\sin{\alpha} - \beta e_y)\Big )^{-1} \Big \{ \beta \gamma e_x e_z, \ e_z (\gamma^{-1} \sin{\alpha} + \cr 
\displaystyle i \cos{\alpha}\sqrt{1 + (\beta \gamma e_x)^2}), \ -\beta \gamma e_x^2 - e_y (\gamma^{-1} \sin{\alpha} + \cr 
\displaystyle i \cos{\alpha} \sqrt{1 + (\beta \gamma e_x)^2} )\Big \} \exp\Big \{i \frac{2 \pi h}{\lambda} (\beta^{-1} \sin{\alpha} - e_y) \Big \} \cr
\displaystyle \times \exp \Big \{- h \cos{\alpha} \frac{2 \pi}{\beta \gamma \lambda} \sqrt{1 + (\beta \gamma e_x)^2}\Big \}. \label{EDR-h}
\end{eqnarray}
One can see, that parameter $h \cos {\alpha} = a$ is the shortest distance between the particle trajectory and the edge of the screen, see Fig.3.
This radiation field leads to the following intensity of DR:
\begin{eqnarray}
&& \displaystyle \frac{d^2 W}{d\omega d\Omega}= \frac{e^2}{\pi^2 c}\frac{1}{4} \Big ((1 + (\beta \gamma e_x)^2)[(\sin{\alpha} - \beta e_y)^2 + \cr 
&& \displaystyle + \cos^2 \alpha (1 - \beta^2 (e_y^2 + e_z^2))] \Big )^{-1} \Big ((e_y^2 + e_z^2) \times \cr 
&& \displaystyle (1 - \beta^2 + \beta^2 \cos^2{\alpha} (1 +  (\gamma e_x)^2)) + (\beta \gamma e_x)^2 \times \cr
&& \displaystyle (e_x^2 + e_z^2) + 2 \beta e_y e_x^2 \sin{\alpha}\Big ) \cr
&& \displaystyle \qquad \qquad \times \exp \Big \{-a \frac{4 \pi}{\beta \gamma \lambda} \sqrt{1 + (\beta \gamma e_x)^2}\Big \}. \label{IDR-h}
\end{eqnarray}
The same expression for the surface current (\ref{ElCurGen}) leads to the following formula for the field of transition radiation on a boundless screen \cite{JETP-2}:
\begin{eqnarray}
&& \displaystyle \bold {E}^{TR}_\infty = \frac{e}{\pi c} \frac{e^{i k r_0}}{r_0}\beta \cos{\alpha} \Big ((\sin{\alpha} - \beta e_y)^2 + \cr 
&& \ \displaystyle + \cos^2{\alpha} (1 - \beta^2 (e_y^2 + e_z^2)) \Big )^{-1}\times \{ e_x e_z, \cr 
&& \quad \displaystyle -\beta e_z  \sin{\alpha}  + e_y e_z, -e_x^2 - e_y^2 + \beta e_y \sin{\alpha} \}. \label{ETR}
\end{eqnarray}
Note that this radiation field leads to the formula for transition radiation intensity derived by Korkhmazyan and Pafomov \cite{JETP-2, Korhmazyan-incident, Korhmazyan, Pafomov}.

By using both of the formulas provided, it is possible to derive an expression for the field of transition radiation, arising when a particle crosses the screen obliquely 
at the distance $h$ to the edge of the screen. This allows to check the transition from TR angular distributions (when $h \rightarrow \infty$) 
to DR ones (when $h \rightarrow 0$). This transition was verified before only for the normal incidence ($\alpha = 0$) in the ultrarelativistic limit \cite{Pap-NIMB98},
while the formulas (\ref{EDR-h}), (\ref{ETR}) have no restrictions on the particle energy and the angle of incidence. 
Transition radiation field generated at the distance $h$ to the edge of the screen is found as a difference between Eqs. (\ref{ETR}) and (\ref{EDR-h}):
\begin{eqnarray}
&& \displaystyle \bold {E}^{TR}_h = \bold {E}^{TR}_{\infty} - \bold {E}^{DR}_h = \frac{e}{\pi c} \frac{e^{i k r_0}}{r_0} \Big ((\sin{\alpha} - \beta e_y)^2 + \cr 
&& \displaystyle + \cos^2{\alpha} (1 - \beta^2 (e_y^2 + e_z^2)) \Big )^{-1}  \Big \{ \beta e_x e_z \Big [ \cos{\alpha} - \cr 
&& \displaystyle - \frac{1}{2} \Big ( \cos{\alpha}  + i \gamma \frac{\sin{\alpha} - \beta e_y}{\sqrt{1 + (\beta \gamma e_x)^2}}\Big ) 
e^{i\varphi} \Big ], \cr 
&& \displaystyle \ e_z \Big [\beta e_y \cos{\alpha} - \beta^2 \sin{\alpha} \cos{\alpha} - \frac{1}{2} \Big (-\cos{\alpha} + \cr 
&& \displaystyle + i \gamma^{-1} \frac{\sin{\alpha}}{\sqrt{1 + (\beta \gamma e_x)^2}}\Big ) 
\times \Big( \sin{\alpha} - \beta e_y - \cr 
&& \displaystyle \qquad - i \gamma^{-1} \cos{\alpha} \sqrt{1 + (\beta \gamma e_x)^2}\Big) e^{i\varphi}\Big ], \cr 
&& \displaystyle \beta \cos{\alpha} (\beta e_y \sin{\alpha} - e_x^2 - e_y^2) + \frac{1}{2} \Big (\gamma^{-1}\cos{\alpha} + \cr 
&& \displaystyle + i \frac{\sin{\alpha} - \beta e_y}{\sqrt{1 + (\beta \gamma e_x)^2}}\Big )\times 
\Big( \beta \gamma e_x^2 + \gamma^{-1}e_y \sin{\alpha} + \cr 
&& \displaystyle \qquad \qquad + i e_y \cos{\alpha} \sqrt{1 + (\beta \gamma e_x)^2}\Big) e^{i\varphi}\Big \}, \label{ETR-h}
\end{eqnarray}
where 
\begin{eqnarray}
\displaystyle \varphi = \frac {2 \pi h}{\lambda} \Big (\beta^{-1} \sin{\alpha} - e_y + i \frac{\cos {\alpha}}{\beta \gamma}\sqrt{1 + (\beta \gamma e_x)^2}\Big ) \label{Phase}
\end{eqnarray}
is denoted. Fig.4 shows the angular distributions of backward TR for various distances to the screen edge $h$. 
As may be seen, the case $h \gg \lambda/2\pi $ gives the well-known lobe-shaped distributions of TR with asymmetry typical for the moderate relativistic energies of the particle.
The opposite condition $h \ll \lambda/2\pi$ results in the one-maximum curve of DR.
\begin{figure}
\center \includegraphics[width=8.50cm, height=6.00cm]{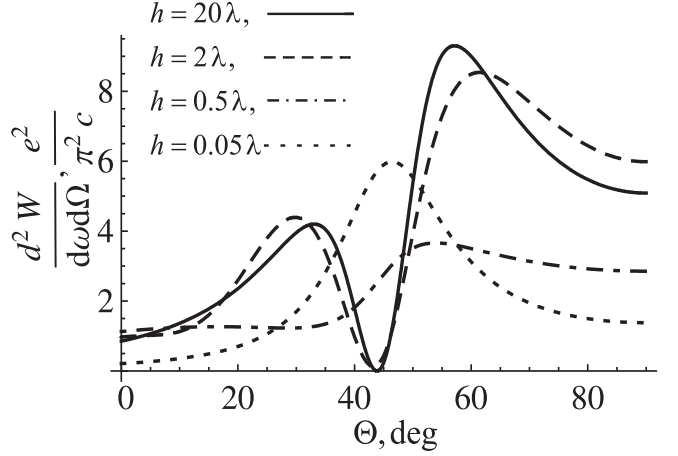}
\center \caption{\label{Fig4}Angular distributions of TR for the various distances from the point of crossing to the edge of the screen.
Parameters: $\gamma = 5, \alpha = \pi/4$, unit radiation vector $\bold {e}_0 = \{\sin{\Theta} \sin{\Phi}, \sin{\Theta} \cos{\Phi}, \cos{\Theta}\}$, 
and $\Phi = 0$}
\end{figure}

The results for DR on a semi-plane may be compared with those obtained via expression for the surface current density without normal component $j_z$ \cite{Kazantsev}. 
The formula for radiation intensity derived in the work cited may be written in our notations as:
\begin{eqnarray}
&& \displaystyle \frac{d^2 W}{d \omega d \Omega} =c\ r^2_0 |\bold {E}^{DR}_h|^2 = \frac{e^2}{\pi^2 c} \frac{\beta}{4} \Big ((e_y^2 + e_z^2) \times \cr 
&& \displaystyle (1 + (\beta \gamma e_x)^2) [(\sin{\alpha} - \beta e_y)^2 + \cos^2{\alpha} \times \cr 
&& \displaystyle (1 - \beta^2 (e_y^2 + e_z^2))] \Big )^{-1} \Big ((1 + (\beta \gamma e_x)^2) \times \cr 
&& \displaystyle (\sqrt{e_y^2 + e_z^2} - e_y) (1 + \beta \sin{\alpha}\sqrt{e_y^2 + e_z^2}) + \cr 
&& \displaystyle (\gamma e_x)^2 (\sqrt{e_y^2 + e_z^2} + e_y) (1 - \beta \sin{\alpha}\sqrt{e_y^2 + e_z^2}) \Big ) \cr 
&& \displaystyle \qquad \times \exp\Big\{-a \frac{4 \pi}{\beta \gamma \lambda}\sqrt{1 + (\beta \gamma e_x)^2}\Big\}. \label{EDR-KS}
\end{eqnarray}
Fig.5 shows the angular distributions of DR for the surface current method being developed (via Eq.(\ref{IDR-h})), 
and the ones according to the formula (\ref{EDR-KS}). As may be seen, the curves noticeably differ from each other in the vicinity of the screen plane: $\Theta \sim \pi/2$,
where the results of Ref.\cite{Kazantsev} become inapplicable due to neglect by the normal component of the current density.
\begin{figure*}
\center \includegraphics[width=14.00cm, height=5.00cm]{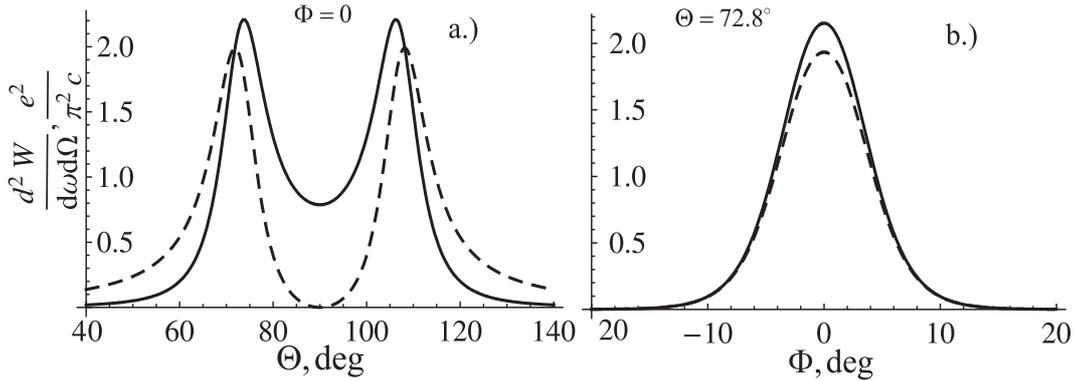}
\center \caption{\label{Fig5}Angular distributions of DR on a semi-plane. 
Solid line - in generalized surface current method: Eq.(\ref{IDR-h}), dashed line - in the ordinary surface current method: Eq.(\ref{EDR-KS}).
Parameters: $\gamma = 10, \lambda = 1$ mm, $a = 2$ mm, $\alpha = 72.8^\circ$}
\end{figure*}

Consider the radiation properties at the distances $\gamma^2 \lambda > r_0 \gg \gamma \lambda/ 2\pi$, i.e. in the so-called pre-wave zone \cite{Verzilov}.
The formula for DR field may be written in the form convenient for the numerical calculations, see Eq.(\ref{ERadSim}):
\begin{eqnarray}
&& \displaystyle E^R_i (\bold r_0, \omega) \approx \frac{i \omega}{2 \pi c}\epsilon_{i j k} {e_0}_j \epsilon_{k l m} n_l A_{m n} 
\cr && \displaystyle \qquad \qquad \times \int \limits_{-\infty}^{\infty}dx \int \limits_{h}^{\infty}dy \
\acute {E}^0_n (\acute {\bold r}, \omega) \frac{e^{i \omega |\bold r_0 - \bold r|/c}}{|\bold r_0 - \bold r|}, \label{Ei}
\end{eqnarray}
where $\epsilon_{i j k}$ is the completely antisymmetric tensor, $A_{m n}$ is the symmetrical rotation matrix: 
\begin{eqnarray}
&& \displaystyle
A_{m n} = \displaystyle\Bigg(\begin{array}{l} 1\quad
\quad 0\qquad \quad 0\cr 0\quad \cos{\alpha}\quad \sin {\alpha}\cr
0\quad \sin{\alpha}\quad -\cos{\alpha}\end{array}\Bigg), \label{M}
\end{eqnarray}
and $\acute {E}^0_n (\acute{\bold r}, \omega)$ is the n-th component of Fourier transform of the particle field  in the frame of reference,
where the charge uniformly moves along the positive direction of $\acute z$-axis (see e.g \cite{Jakson, JETP-2}),
and as usual $\acute r_i = A_{i j} r_j$. 

Integrals (\ref{Ei}) are evaluated numerically, Wolfram Mathematica 6.0 is used. The results for different angles of incidence of a particle having the moderate relativistic energy are presented in Fig.6.
One can see that as the distance to observer decreases, the maximum in the angular distributions moves to the plane of the screen.
It is essential that for the large values of the incidence angle $\alpha$, the radiation registered in pre-wave zone is concentrated in the vicinity of the screen plane, see Fig.6 b.). 
As the total energy losses don't depend upon a distance $r_0$, the narrowing of the angular distributions in the plane $\Phi = 0$
is compensated by broadening of the ones in perpendicular plane. Furthermore, the total radiation losses depend neither upon the angle of incidence $\alpha$ 
nor upon a method they are calculated with: the generalized surface current method, or the one of the classical optics \cite{Kazantsev, Bolotovsky-66, VDB-DR}
(see in more detail \cite{JETP-2}). The last fact just means that the total energy radiated by an electron is determined only by its energy and a distance to the screen edge:
\begin{eqnarray}
W = \frac{3}{8}\frac{e^2 \beta^2}{a} \gamma \label{W}
\end{eqnarray}
Finally, note that in relativistic case the radiation losses don't depend upon a screen shape, for example, on the corner angle of a wedge \cite{VDB-DR}.
\begin{figure*}
\center \includegraphics[width=14.50cm, height=5.50cm]{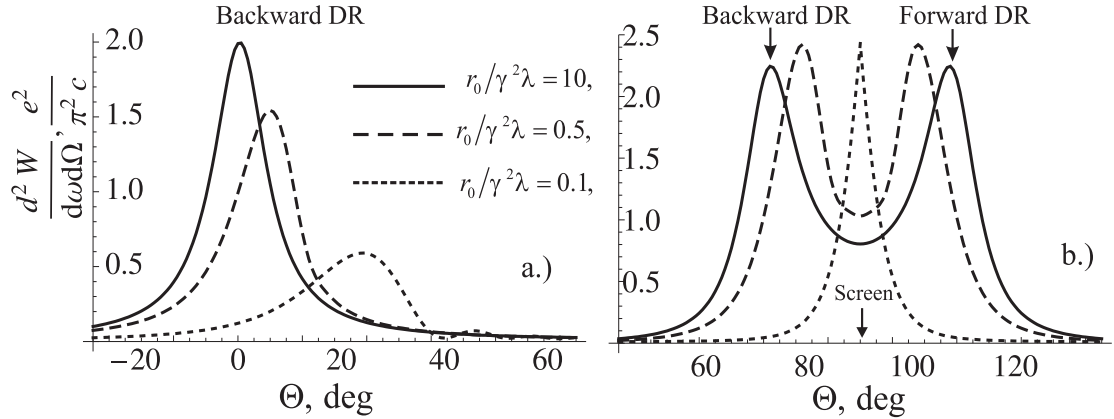}
\center \caption{\label{Fig6}Angular distributions of DR on a semi-plane for various distances to the observer. 
Parameters: $\gamma = 10, \lambda = 1$ mm, $a = 2$ mm, $\Phi = 0$. a.) $\alpha = 0$, b.) $\alpha = 72.8^\circ$}
\end{figure*}

\section{DR of a particle on a slit in cylinder}
\label{Sect5}

Focusing of diffraction radiation and transition radiation by concave surfaces is a good variant to suppress pre-wave zone effect 
for the use of DR and TR in beam diagnostics \cite{Karataev-PLA, Rezaev}. 
TR generated by an ultrarelativistic particle crossing an ideally-conducting concave screen was considered for the first time in the paper \cite{Tilinin}. 
More rigorous theory for this case was developed recently \cite{JETP}. 
Here, the case of DR generated by a particle moving obliquely through a slit of width $b$ in cylinder (see Fig.7) is studied
via the generalized surface current method. 
For this purpose, one should make the appropriate modifications in the formula for radiation field (\ref{Ei}). 
The component of the particle field $\acute{E}^0_n (A_{i j} r_j, \omega)$ now is expressed through cylindrical coordinates: 
\begin{eqnarray}
\bold r = \{x, \rho \sin{\phi}, \rho (1 - \cos{\phi})\}.\label{cylcoord}
\end{eqnarray}
The element of integration on cylinder is $dS = \rho dx d\phi$, 
and the limits are $(-\infty, \infty)$ for $x$ and 
\begin{eqnarray}
&& [-\arcsin{(r_0/\rho)}, -\arcsin{(b/2\rho)}] \cr && \qquad \qquad \cup \ [{\arcsin{(b/2\rho)}, \arcsin{(r_0/\rho)}}]\label{limits}
\end{eqnarray}
for $\phi$. 
Unit normal has the following components:
\begin{eqnarray}
\bold n = \{0, -\sin{\phi}, \cos{\phi}\}.\label{normal}
\end{eqnarray}
\begin{figure}
\center \includegraphics[width=4.50cm, height=5.50cm]{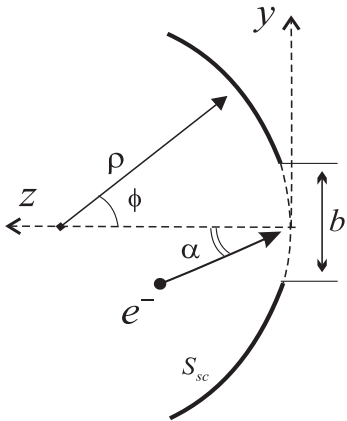}
\center \caption{\label{Fig7}Generation of DR by passing of a particle through a slit in cylinder}
\end{figure}

We study backward DR of a relativistic particle in the pre-wave zone: $r_0 \ll \gamma^2 \lambda$. 
In this case, it is convenient to use the so-called projection angles $\Theta_x, \Theta_y$ (see e.g. \cite{Pap-NIMB98, JETP})
counted from the specular reflection direction: $\Phi = 0, \Theta = \alpha$. 
Firstly, let us make sure that in the case of the large cylinder radius $\rho \gg r_0$ we get the case of DR on a slit in the flat screen \cite{JETP-2}.
As one can see in Fig.8 a.), the line for the zero-width slit completely coincides with that for TR. 
The analytical transition from the formulas of DR to the ones of TR, taking a place when $b=0$, for the flat screen was also made in the paper cited.

If the cylinder curvature is not so large compared to the distance $r_0$, the focusing effect for DR may be observed.
For instance, DR focusing for a definite value of the slit width and various radii of curvature is shown in Fig.8 b.).
Note that asymmetry in the angular distributions vanishes for the case $\rho \sim r_0$, while the opposite dependence was predicted for TR using another method \cite{JETP}.

\section{Discussion}
\label{Sect6}

We have developed the surface current method in the macroscopic theory of TR and DR, which may be considered as a generalization of the classical Fock's method in the diffraction theory.
By using this method we have found several exact solutions of DR problems in the wave zone, studied the features of DR from a semi-plane in the pre-wave zone, 
and the ones of DR generated on a slit in cylinder where the radiation focusing is observed. The exact transition between TR and DR phenomena for the arbitrary particle energy 
and the arbitrary incidence angle has been demonstrated as for a slit, as for a semi-plane.
\begin{figure*}
\center \includegraphics[width=15.50cm, height=5.50cm]{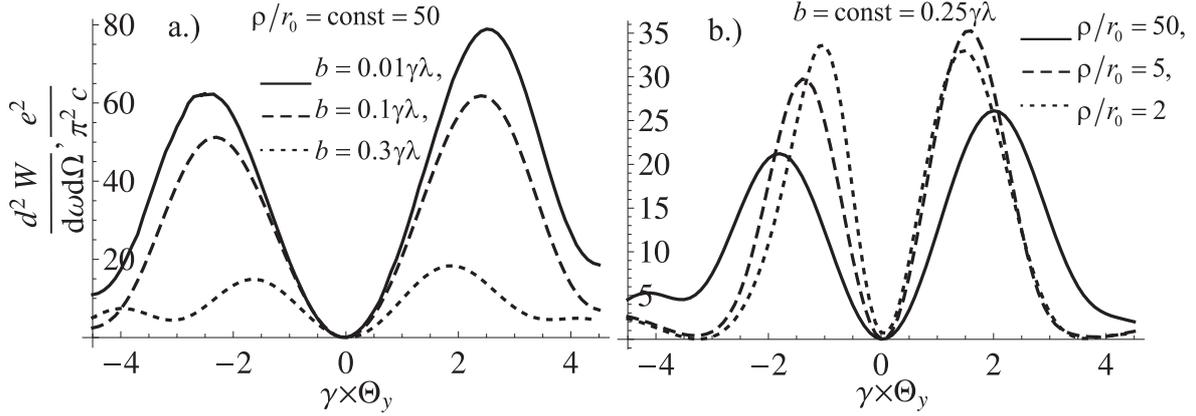}
\center \caption{\label{Fig8}Angular distributions of DR on a slit in cylinder: a.) for a constant value of the surface curvature, b.) for a constant value of the slit width.
Parameters: $\gamma = 20, \lambda = 1$ mm, $\alpha = \pi/4$, $r_0/\gamma^2 \lambda = 0.1$,
unit radiation vector $\bold {e}_0 = \{\sin{\Theta_x}, \cos{\Theta_x} \sin{(\alpha + \Theta_y)}, \cos{\Theta_x} \cos{(\alpha + \Theta_y)}\}$ and $\Theta_x = 0$}
\end{figure*}

In conclusion, let us discuss the limits of applicability for the method developed.
Firstly, only the ideal conductivity leads to the non-zero value of TR intensity in a plane of the screen \cite{Landau-8}.
Accordingly, one can expect that for all real conductors intensity of DR on a semi-plane or grating would also vanish in this direction.
Secondly, we consider the \textit{macroscopic} theory whose limits of applicability were investigated for TR in Ref.\cite{Ryazanov-JETPLetters}.
By using the results of the paper cited, one can conclude that TR theory developed is not suitable for the case of non-relativistic particles: $\beta \ll 1$,
and of the grazing incidence: $\alpha \sim \pi/2$. The same conditions hold true for DR as well. 
However, it may be realized a case when a fast particle moves parallel to the plane of the screen: $\alpha = \pi/2$,
but the macroscopic approach is valid. It is so, as long as the distance to the screen is much larger 
than the effective dimension of the spatial averaging: $a \gg \lambda/2\pi$.
Finally, the region of frequencies where the approximation of ideal conductivity is valid is bounded above by the optical ones.

\

{\bf Acknowledgments}

\
We are grateful to Prof. M.I. Ryazanov and Dr. A.A. Tishchenko for stimulating criticism and fruitful discussions. 
The work is partially supported by the grant of Russian Foundation for Basic Research No.08-02-09506-mob-z.








\

\end{document}